\documentstyle{elsart}
\begin{document}
\begin{frontmatter}

\title{Nonextensive distribution and factorization of the joint probability}

\author[Ism]{Qiuping A. Wang,}
\author[Univ]{Michel Pezeril,}
\author[Ism]{Laurent Nivanen,}
\author[Ism]{and Alain Le M\'{e}haut\'{e}}
\address[Ism]{Institut Sup\'{e}rieur des Mat\'{e}riaux du Mans, 44, Av.
Bartholdi, 72000 Le Mans, France}
\address[Univ]{Laboratoire de Physique de l'\'etat Condens\'e,
Universit\'e du Maine,\\
72000 Le Mans, France}

\begin{abstract}
The problem of factorization of a nonextensive probability
distribution is discussed. It is shown that the correlation
energy between the correlated subsystems in the canonical
composite system can not be neglected even in the thermodynamic
limit. In consequence, the factorization approximation should be
employed carefully according to different systems. It is also
shown that the zeroth law of thermodynamics can be established
within the framework of the Incomplete Statistical Mechanics
($ISM$).

\end{abstract}


{\em PACS index codes\/} : 02.50.-r, 05.20.-y, 05.30.-d,05.70.-a
\end{frontmatter}

\newpage
\section{Introduction}
The nonextensive probability distribution

\begin{equation}
p_i=\frac{[1-(1-q)\beta(E_i-C)]^\frac{1}{1-q}}{Z_q}
                                                        \label{1}
\end{equation}
plays a decisive role for the success of the generalized
statistical mechanics \cite{Tsal98,Wang00} because it is capable
of reproducing unusual distributions of non gaussian type which
are met frequently in nature (see reference \cite{Tsal98} and the
references there-in). Very recently, distribution functions of
type Eq. (\ref{1}) were recognized by the so-called
eigencoordinates method with high level of authenticity as good
description of the amplitude distribution of earthquake noises
which are proved to be fractal and strongly correlated
\cite{Raoul}. In Eq. (\ref{1}), $\beta$ is the generalized
inverse temperature, $E_i$ is the energy of the system in the
state $i$, $C$ a constant to assure the invariance of the
distribution through uniform translation of energy spectrum $E_i$,
and $Z_q$ is given by
\begin{equation}                                        \label{2}
Z_q=\sum_i[1-(1-q)\beta (E_i-U_q)]^\frac{1}{1-q}
\end{equation}
or by
\begin{equation}                                        \label{2a}
Z_q=\sum_i[1-(1-q)\beta (E_i-U_q)]^\frac{q}{1-q} , \cite{Comment}
\end{equation}
in Tsallis' multi-fractal inspired scenario \cite{Tsal98} with
$U_q$ the internal energy of the system and
$\beta=\frac{Z^{q-1}}{kT}$, and by
\begin{equation}                                        \label{3}
Z_q=\left[\sum_i[1-(1-q)\beta E_i]^\frac{q}
{1-q}\right]^\frac{1}{q}.
\end{equation}
with $\beta=\frac{Z^{1-q}}{kT}$ in Wang's {\it incomplete
statistics} scenario \cite{Wang00} devoted to describe inexact or
incomplete probability distribution \cite{Reny66} due to neglected
interactions in the system hamiltonian. It is noteworthy that Eq.
(\ref{1}) is a canonical distribution function for isolated
system in terms of its total energy $E_i$.

In this paper, we discuss the factorization problem of Eq.
(\ref{1}) and some of its consequences when the canonical system
is composed of correlated subsystems or particles. At the same
time, we also comment on some interesting applications of Eq.
(\ref{1}) found in the literature.

\section{Factorization approximation}

In Boltzmann-Gibbs-Shannon statistics (BGS), thanks to the easy
factorization of the exponential distribution functions, i.e.,
\begin{equation}                                        \label{4}
\sum_ie^{-\beta\sum_je_{ij}}=\prod_j\sum_ie^{-\beta e_{ij}}.
\end{equation}
the system distribution function can yield one particle one in
terms of the single body energy $e_{ij}$ in the case of
independent particles or of mean-field method with
$E_i=\sum_{j=1}^N e_{ij}$, where $j$ is the index and N the total
number of the distinguishable particles.

But this approach is impossible in the case of the nonextensive
distribution, because,

\begin{equation}                                        \label{5}
\sum_i[1-(1-q)\beta\sum_je_{ij}]^\frac{1}{1-q}\neq
\prod_j\sum_i[1-(1-q)\beta e_{ij}]^\frac{1}{1-q}.
\end{equation}

This inequality makes it difficult to apply the nonextensive
distribution even to systems of independent particles, because the
total partition function can not be factorized into single
particle one. So in the definition of the total system entropy
$S_q=f[p(E_i),q]$, $p(E_i)$ must be the probability of a
microstate of the system and $E_i$ can not be replaced by
one-body energy $e_{ij}$. However, in the literature, we find
applications of Tsallis' distribution in which $E_i$ is
systematically replaced by $e_{ij}$ without explanation
\cite{Plas93,Bog96,Kan96,Ante97,Wang98,Lava98}. The first
examples\cite{Plas93,Kan96} are related to the polytropic model
of galaxies and the authors have taken the one-body energy of
stars and of solar neutrinos ($\epsilon=\Psi+\frac{v^2}{2}$) as
$E_i$. Other examples are the peculiar velocity of galaxy
clusters ($e \sim v^2$)\cite{Lava98} and the electron plasma
turbulence where the electron single site density $n(r)$ was
taken as system (electron plasma) distribution function and the
total energy was calculated with the one-electron potential
$\phi(r)$ \cite{Bog96,Ante97}. The last case is the application of
nonextensive blackbody distribution to laser systems where the
atomic energy levels were taken as laser system energy in
equation (8) of reference \cite{Wang98}. As a matter of fact, in
above examples, the calculated entropies and distributions are
one-particle ones, as we always do in BGS framework. Consequently,
considering Eq. (\ref{5}), they are only approximate applications
of the $exact$ Tsallis' distribution.

The legitimacy of the above mentioned applications depends on the
approximation with which we write Eq. (\ref{5}) as an equality.
One of the solutions is the limit of $q\rightarrow 1$ where Eq.
(\ref{5}) tends to an equality. But this solution does not hold
for the cases of $q$ value very different from unity, as in the
above mentioned examples of applications. Another way out is the
{\it factorization approximation} proposed in order to obtain the
nonextensive Fermi-Dirac and Bose-Einstein distributions
\cite{Buyu93} which read
\begin{equation}                                    \label{6}
\langle
n_q\rangle=\frac{g}{[1+(q-1)\beta(e-\mu)]^\frac{1}{q-1}\pm1}
\end{equation}
where $g$ is the degeneracy of the level with energy $e$. As in
the case of the corrected Boltzmann distribution (i.e.
$\frac{n_q}{g}\ll 1$), Eq. (\ref{6}) can be reduced to

\begin{equation}                                      \label{7}
p(e)\sim\frac{n_q}{g}=[1-(1-q)\beta(e-\mu)]^\frac{1}{1-q}
\end{equation}

where $p(e)$ is one-particle probability and $e$ one-particle
energy. So we can say that only in this approximation the above
mentioned applications are justified.

It is worth mentioning that, although approximate, the above
successful applications were the first proofs of the existence of
Tsallis type distributions (Eq. (\ref{1})) in nature. In
addition, this is a parametrized distribution. So in many cases,
what we neglect in the approximation may be compensated (at least
partially) by a different value of $q$ fixed empirically. The
interest of the applications mentioned above is to show that this
kind of nonextensive (nonadditive) probability can really
describe some non-gaussian type peculiar distributions we
observe. That is what is important in practice.

Nevertheless, an approximation has sometimes in itself theoretical
importance when it concerns the basic foundation of a theory.
That is the case of the factorization approximation.

\section{Factorization of the joint probability}
The factorization approximation is a forced marriage between the
right-hand side and the left-hand side of Eq. (\ref{5}), that is
we write just like that :
\begin{equation}                                        \label{8}
[1-(1-q)\beta\sum_je_j]^\frac{1}{1-q}\simeq \prod_j[1-(1-q)\beta e
_j]^\frac{1}{1-q}.
\end{equation}
where we keep only the index $j$ of the particles, as we do from
now on in this section. What is neglected in this approximation is
the difference $\Delta$ between the right-hand side and the
left-hand side of Eq. (\ref{8}) which has been investigated in
reference \cite{Wang97} with a two-level system for the simple
case where $q>1$ (or $q<1$), $e_j>0$ (or $e_j<0$) and $\mu=0$.
Under these harsh conditions, $\Delta$ turns out to be very small
at normal temperatures for mesoscopic or macroscopic systems (with
important particle number $N$). But it is not the case for $q<1$
and $e_{j}>0$ with in addition $\mu\neq 1$. So in general, we can
not write Eq. (\ref{8}).

Eqs. (\ref{5}) and Eq. (\ref{8}) can be discussed in another way
as follows. If we replace $\sum_je_j$ by the total energy $E$ in
Eq. (\ref{8}), we get :

\begin{equation}                                        \label{8a}
[1-(1-q)\beta E]^\frac{1}{1-q}=\prod_j^N[1-(1-q)\beta
e_j]^\frac{1}{1-q}.
\end{equation}

for $N$ subsystems (or particles with energy $e_j$ where
$j=1,2,...N$) of a composite system with total energy $E$ at a
given state. This is just the factorization of the joint
probability $p(E)$ as a product of all $p(e_j)$ :
\begin{equation}                                    \label{9}
p(E)=\prod_{j=1}^N p(e_j).
\end{equation}

With Eq. (\ref{8a}) or (\ref{9}), strictly speaking, we can not
write
\begin{equation}                                    \label{9a}
E=\sum_je_j.
\end{equation}
But in Tsallis' scenario, Eq. (\ref{9a}) is necessary for
establishing the zeroth law of thermodynamics \cite{Abe99}. Abe
studied this problem with ideal gas model and concluded that Eq.
(\ref{9a}) can hold for $e_j>0$, $0<q<1$ and $N\rightarrow
\infty$. That is the correlation energy $E_c=E-\sum_je_j$ between
the maybe strongly correlated subsystems or particles can be
neglected. But this is of course not a general conclusion for any
$q$ value or any system. In what follows, we will try to give the
general expression of the correlation energy in $ISM$ because
this relation is implicit in Tsallis' scenario \cite{Corr}. The
following discussion is for $0<q<\infty$, the permitted interval
of $q$ value in $ISM$ \cite{Wang00}.

If $N=1$, from Eq. (\ref{9}), we naturally obtain $E=e_1$ so
$E_c=0$.

If $N=2$, we obtain :
\begin{eqnarray}                                    \label{10}
aE  & = & (1+ae_1)(1+ae_2)-1 \\ \nonumber
    & = & ae_1+ae_2+a^2e_1e_2=a\sum_{i=1}^2 e_i+a^2e_1e_2
\end{eqnarray}
where $a=(q-1)\beta$. So $E_c=ae_1e_2$.

If $N=3$, we get :
\begin{eqnarray}                                    \label{11}
aE   & = & (1+ae_1)(1+ae_2)(1+ae_3)-1 \\ \nonumber
 & = & a(e_1+e_2+e_3)+a^2(e_1e_2+e_1e_3+e_2e_3)+a^3(e_1e_2e_3)\\
\nonumber
  & = & a\sum_{i=1}^3 e_i+a^2\sum_{i_1<i_2}^{3(\frac{3!}{2!1!}
  terms)}e_{i_1}e_{i_2}+a^3\prod_{i=1}^3 e_i.
\end{eqnarray}
and
\begin{eqnarray}                                    \label{11a}
E_c & = & a(e_1e_2+e_1e_3+e_2e_3)+a^2(e_1e_2e_3).
\end{eqnarray}
 When $N\rightarrow\infty$, this is a infinite product problem. In
general, we obtain :

\begin{eqnarray}                                   \label{12}
aE   & = & \prod_{i=1}^N (1+ae_i)-1   \\ \nonumber
 & = & a\sum_{i=1}^N e_i+a^2\sum_{i_1<i_2}
 ^{N(\frac{N!}{2!(N-2)!}
  terms)}e_{i_1}e_{i_2} \\ \nonumber
  & + & a^3 \sum_{i_1<i_2<i_3}
  ^{N(\frac{N!}{3!(N-3)!}
  terms)}e_{i_1}e_{i_2}e_{i_3}+ ... +a^N\sum_{i_1<i_2<...<i_N}^{N(\frac{N!}{N!(N-N)!}
  terms)}\prod_{i=1}^N e_i \\ \nonumber
 & = & \sum_{k=1}^Na^k\sum_{i_1<i_2<...<i_k}
 ^{N(\frac{N!}{k!(N-k)!}
  terms)}\prod_{j=1}^k e_{i_j}  \\ \nonumber
 & = & aE_0+aE_c
\end{eqnarray}
where $E_0$ is the system energy given by Eq. (\ref{9a}) for
independent subsystems like the particles of perfect gas. $E_c$
is given by :

\begin{eqnarray}                                    \label{14}
E_c & = & \prod_{i=1}^N (1+ae_i)-E_0     \\ \nonumber & = &
\sum_{k=2}^Na^{k-1}\sum_{i_1<i_2<...<i_k}
^{N(\frac{N!}{k!(N-k)!}terms)}\prod_{j=1}^k e_{i_j}
\end{eqnarray}

The value of $E_c$ is in general difficult to estimate. We can
discuss it in the following way.

For $q>1$ (or $0<q<1$) and $e_i>0$ (or $e_i<0$), $E_c$ may be very
big when $N\rightarrow\infty$ because $\sum_i(ae_i)$ may diverge.
So it is in general not negligible.

For $q>1$ (or $0<q<1$) and $e_i<0$ (or $e_i>0$),
\begin{eqnarray}                                    \label{15}
E_c & = &
\sum_{k=2}^N(-1)^{k-1}[(1-q)\beta]^{k-1}\sum_{i_1<i_2<...<i_k}^{N(\frac{N!}{k!(N-k)!}
terms)}\prod_{j=1}^k |e_{i_j}|
\end{eqnarray}
The sign of each term varies alternatively, which may turn out to
cancel $E_c$ when $N\rightarrow\infty$. But this is only a
possibility. In general, we can not say that $E_c$ is negligible.

Considering Eq. (\ref{9}), the meanvalue of $E_c$ is given by
following integral in phase-space $\Omega (\tau)$ :

\begin{eqnarray}                              \label{16}
U_c & = & \int d\tau p(E)E_c   \\ \nonumber
    & = & \sum_{k=2}^N(-1)^{k-1}[(1-q)\beta]^{k-1}
    \sum_{i_1<i_2<...<i_k}^{N(\frac{N!}{k!(N-k)!}
    terms)}\prod_{j=1}^k |\int d\tau p(e_{i_j})e_{i_j}|  \\ \nonumber
    & = & \sum_{k=2}^N(-1)^{k-1}[(1-q)\beta]^{k-1}
    \sum_{i_1<i_2<...<i_k}^{N(\frac{N!}{k!(N-k)!}
    terms)}\prod_{j=1}^k |U_{i_j}|
\end{eqnarray}

The conclusion of this section is that $U_c$ is in general not
negligible. So one must be very careful in using nonextensive
distribution for correlated subsystems within factorization
approximation, especially in the case of the nonextensive quantum
statistics derived on the basis of this approximation
\cite{Buyu93}.

\section{Re-establishment of the zeroth law of thermodynamics}

In this section, we will discuss an application of the so called
Incomplete Statistical Mechanics ($ISM$) \cite{Wang00}, a new
version of the nonextensive statistical mechanics ($NSM$) based
on the normalization condition $\sum_ip_i^q=1$ where $q$ is a
positive parameter.  $q\neq 1$ corresponds to the fact that the
probability distributions $\{ p_i \}$ with {\it incomplete random
variables} \cite{Reny66} do not sum to one. If $q=1$, the random
variables of the problem become complete and we recover the
conventional normalization ($CN$) $\sum_ip_i=1$. $ISM$ was
proposed as a consequence of the study of the fundamental
theoretical problems \cite{Tsal98,Comment} of Tsallis version of
$NSM$ with escort probability \cite{Tsal98}. The fundamental
philosophy of $ISM$ is that $CN$ is difficult to be applied to
the systems with important complicated interactions or
correlations because, in this case, our "incomplete" knowledge
about the states of the systems does not permit us to sum all the
(exact) probabilities. With simpler systems, this human ignorance
could be neglected and $CN$ holds.

In our opinion, $ISM$ is not just another form of Tsallis
nonextensive statistics, contrary to what some scientists thought.
Some of the significant differences is discussed in
Ref.\cite{Comment,Corr}. In what follows, we will re-establish, in
a precise way, the zeroth law which was established within Tsallis
theory only in the case of factorization approximation in
neglecting the correlation energy.

In incomplete statistical mechanics, we have following equations
for a composite system containing two correlated subsystems $A$
and $B$  :

\begin{equation}                                    \label{4a}
S_q(A+B)=S_q(A)+S_q(B)+\frac{q-1}{k}S_q(A)S_q(B),
\end{equation}

\begin{equation}                                    \label{8b}
E_{ij}(A+B)=E_i(A)+E_j(B)+(q-1)\beta E_i(A)E_j(B),
\end{equation}

\begin{equation}                                    \label{9b}
U_q(A+B)=U_q(A)+U_q(B)+(q-1)\beta U_q(A)U_q(B)
\end{equation}

\begin{equation}
S_q=k\frac{Z_q^{q-1}-1}{q-1}+k\beta Z_q^{q-1}U_q,    \label{34}
\end{equation}
and
\begin{equation}                                    \label{35}
\beta=\frac{Z_q^{1-q}}{kT}.
\end{equation}
In Eq. (\ref{4a}), there is a plus sign "+" before the correlated
term. But it was minus sign "-" in Tsallis formalism. Eqs.
(\ref{8b}) and (\ref{9b}) show the same $q$-dependence of the
correlation terms as Eq. (\ref{4a}). They do not exist in Tsallis
formalism with escort probability \cite{Corr}. The $q$-dependence
of the above equations is of crucial importance for the
establishment of the zeroth law.

From Eq. (\ref{4a}), a small variation of the total entropy can
be written as :

\begin{eqnarray}                                    \label{37}
\delta S_q(A+B) & = & [1+\frac{q-1}{k}S_q(B)]\delta S_q(A)+
[1+\frac{q-1}{k}S_q(A)]\delta S_q(B) \\ \nonumber
    & = & [1+\frac{q-1}{k}S_q(B)]
\frac{\partial S_q(A)}{\partial U_q(A)} \delta U_q(A)
\\ \nonumber
    & + & [1+\frac{q-1}{k}S_q(A)] \frac{\partial
S_q(B)}{\partial U_q(B)} \delta U_q(B).
\end{eqnarray}

And from Eq. (\ref{9b}), the variation of the total internal
energy is given by :

\begin{eqnarray}                                    \label{38}
\delta U_q(A+B) & = & [1+\frac{q-1}{k}U_q(B)]\delta U_q(A)+
[1+\frac{q-1}{k}U_q(A)]\delta U_q(B).
\end{eqnarray}
It is supposed that the total system $(A+B)$ is completely
isolated. So $\delta U_q(A+B)=0$ which leads to :
\begin{eqnarray}                                    \label{39}
\frac{\delta U_q(A)}{1+\frac{q-1}{k}U_q(A)}= -\frac{\delta
U_q(B)}{1+\frac{q-1}{k}U_q(B)}
\end{eqnarray}

When the composite system $(A+B)$ is in $equilibrium$, $\delta
S_q(A+B)=0$. In this case, Eqs. (\ref{37}) and (\ref{39}) lead us
to :

\begin{eqnarray}                                    \label{40}
\frac{1+\frac{q-1}{k}U_q(A)}{1+\frac{q-1}{k}S_q(A)}\frac{\partial
S_q(A)}{\partial U_q(A)}=
\frac{1+\frac{q-1}{k}U_q(B)}{1+\frac{q-1}{k}S_q(B)}\frac{\partial
S_q(B)}{\partial U_q(B)}.
\end{eqnarray}
With the help of Eqs. (\ref{34}) and (\ref{35}), it is
straightforward to show that, in general :

\begin{eqnarray}                                    \label{41}
\frac{1+\frac{q-1}{k}U_q}{1+\frac{q-1}{k}S_q}=Z_q^{1-q}
\end{eqnarray}
which recasts Eq. (\ref{40}) as follows :
\begin{eqnarray}                                    \label{42}
Z_q^{1-q}(A)\frac{\partial S_q(A)}{\partial U_q(A)}=
Z_q^{1-q}(B)\frac{\partial S_q(B)}{\partial U_q(B)}
\end{eqnarray}
or
\begin{eqnarray}                                    \label{43}
\beta (A)=\beta (B)
\end{eqnarray}
where $\beta$ is the generalized inverse temperature defined in
Eq. (\ref{35}). Eq. (\ref{42}) or (\ref{43}) is the generalized
zeroth law of thermodynamics which describes the thermodynamic
relations between different nonextensive systems in thermal
equilibrium.

one of the important meanings of Eq. (\ref{43}) is that the
thermal equilibrium of a system is now caracterized by $\beta$
but not $T$. This allows us to have an explicit distribution with
Eq. (\ref{1}), which becomes implicit in Tsallis formalism with
Eq. (\ref{2}) and (\ref{2a}).

\section{Conclusion}
Our conclusion is that all applications of Tsallis' distribution
relative to correlated subsystems or particles should be
considered to be valid only under the {\it factorization
approximation} and the condition $\frac{n_q}{g}\ll 1$. This
factorization approximation must be employed carefully because
the correlation energy between the correlated subsystems is in
general not negligible even in the thermodynamic limit
($N\rightarrow\infty$). The zeroth law of thermodynamics can be
established in the framework of {\it Incomplete Statistical
Mechanics} without any approximation.

\section{Acknowledgments}

We acknowledge with great pleasure the very useful discussions
with Professor Sumiyoshi Abe on some points of this work.


\begin{thebibliography}{99}
\bibitem {Corr}
For a composite system containing two correlated subsystems $A$
and $B$, from the factorization of joint probability
$p_{ij}(A+B)=p_i(A)p_j(B)$ in Tsallis scenario with escort
probability, we have :
$\frac{E_{ij}(A+B)-U_q(A+B)}{\sum_{ij}p_{ij}^q(A+B)}=
\frac{E_i(A)-U_q(A)}{\sum_ip_{i}^q(A)}+
\frac{E_j(B)-U_q(B)}{\sum_jp_j^q(B)}+ \frac{(q-1)\beta
[E_i(A)-U_q(A)][E_j(B)-U_q(B)]}{\sum_ip_i^q(A)\sum_jp_j^q(B)}$.
This equation makes it impossible to find the correlation energy
$E_c(A,B)$ and its meanvalue.
\end{thebibliography}
\end{document}